\documentclass[aps,prd,showpacs,twocolumn,superscriptaddress,amsmath,amsfont,amssymb]{revtex4}
\pdfoutput=1
\usepackage{color,graphicx}
\usepackage{color}
\usepackage{slashed}
\usepackage[english]{babel}

\definecolor{Red}{rgb}{1.,0.,0.}

\begin{document}

\title{Light colored scalars from grand unification and the forward-backward asymmetry in $ t\bar t$ production}

\author{Ilja Dor\v sner} 
\email[Electronic address:]{ilja.dorsner@ijs.si}
\affiliation{Department of Physics, University of Sarajevo, Zmaja do Bosne 33-35,
71000 Sarajevo,
Bosnia and Herzegovina}
\affiliation{}

\author{Svjetlana Fajfer} 
\email[Electronic address:]{svjetlana.fajfer@ijs.si} \affiliation{Department of Physics,
  University of Ljubljana, Jadranska 19, 1000 Ljubljana, Slovenia}
\affiliation{J. Stefan Institute, Jamova 39, P. O. Box 3000, 1001 Ljubljana, Slovenia}

\author{Jernej F. Kamenik} 
\email[Electronic address:]{jernej.kamenik@ijs.si} 
\affiliation{J. Stefan Institute, Jamova 39, P. O. Box 3000, 1001
  Ljubljana, Slovenia}

\author{Nejc Ko\v{s}nik} 
\email[Electronic address:]{nejc.kosnik@ijs.si} 
\affiliation{J. Stefan Institute, Jamova 39, P. O. Box 3000, 1001
  Ljubljana, Slovenia}

\date{\today}

\begin{abstract}
The experimental results on the $ t \bar t$ production cross section at the Tevatron are well described by the QCD contributions within the standard model, while the recently measured forward-backward asymmetry is larger than predicted within this framework.  We consider light colored scalars  appearing in a particular $SU(5)$ GUT model within the 45-dimensional Higgs representation.
A virtue of the model is that it connects the presence of a light colored $SU(2)$ singlet ($\Delta_6$) and a color octet weak doublet  ($\Delta_1$) with bounds on the proton lifetime, which  constrain the parameter space of both scalars.  We find that both the total $t \bar t$ production cross section and the forward-backward asymmetry can be accommodated simultaneously within this model. The experimental results  prefer a region for the mass of  $\Delta_6$ around $400$ GeV, while $\Delta_1$ is then constrained to have a mass around the TeV scale as well.  We analyze possible experimental signatures and find that $\Delta_6$ associated top production could be probed in the $t\bar t$+jets final states at Tevatron and the LHC.
\end{abstract}


\maketitle

\section{Introduction}

The CDF and D0 experiments at the Tevatron have produced many important and useful results in top quark physics. Recently the measurements of the forward-backward asymmetry in $t\bar t$ production (FBA) \cite{FBAExp,FBAExpCDF} have attracted a lot of attention due to the more than $2 \sigma$ discrepancy of the most precise experimental result \cite{FBAExpCDF} compared to the ${\cal O}(\alpha_s^3)$ interference effect predicted within the standard model (SM) \cite{AFBtheo}.  At the same time, other CDF and D0 results on  top quark properties and processes exhibit a very good agreement with SM predictions \cite{topExp}.  Several scenarios of new physics (NP) have already been considered, trying to explain this discrepancy \cite{Djouadi:2009nb,Cheung:2009ch,Jung:2009jz,Frampton:2009rk}.  


Among possible NP proposals, the path of gauge unification in the form of grand unified theories (GUTs) is perhaps one of the most appealing. Traditionally, the two main challenges of GUTs have been achieving SM gauge couplings' unification and ensuring stability of the proton beyond present experimental limits.  Recently, non-supersymmetric SU(5) GUT models~\cite{Dorsner:2006dj,Dorsner:2007fy,Perez:2007rm} that incorporate 45-dimensional scalar representation have been found to satisfy the first criterion. An important observation was that some of the $45$ states can be relatively light and play an interesting role in ensuring sufficient proton stability.  With this in mind, we have recently investigated the role of  scalar leptoquarks also appearing within such a representation in charm meson decays \cite{Dorsner:2009cu}.  

In the present study we investigate a colored $SU(2)_L$ singlet with electric charge $4/3$ ($\Delta_6$), which is incorporated in the $45$-dimensional representation of $SU(5)$, as a possible explanation of the FBA discrepancy. Direct experimental constraints on the mass and couplings of the new state come from flavour and collider experiments. In addition however, they are constrained indirectly due to gauge couplings' unification and with the proton not decaying too fast. In particular, it turns out that the later two conditions require another relatively light state $(\Delta_1)$, which is an octet of color, doublet of $SU(2)_L$ and has hypercharge one half.



The paper is organized in the following sections: In section II we introduce the candidate $\Delta_{1,6}$ states within a GUT model and discuss constraints coming from gauge couplings' unification and proton decay. In section III we discuss the phenomenology of the two $\Delta$'s in hadronic $t\bar t$ production at the Tevatron, while other related phenomenology and constraints are discussed in section IV. We briefly cover possible search strategies for $\Delta_6$ at the Tevatron and the LHC in section V, before summarizing our results in section VI.

\section{$\Delta_1$ and $\Delta_6$}

The scalar fields we discuss---$(\bm{}\overline{\bm{3}},\bm{1}, 4/3)$ and $(\bm{8},\bm{2}, 1/2)$---naturally emerge in a theoretically well-motivated class of grand unified scenarios. Namely, they both reside in a $45$-dimensional scalar representation of $SU(5)$. And, this representation should be a part of any simple renormalizable scenario without supersymmetry (SUSY) together with a $5$-dimensional scalar representation---$\bm{5} \equiv (\Psi_D,
\Psi_T) = (\bm{1},\bm{2},1/2)\oplus(\bm{3},\bm{1},-1/3)$---to generate viable masses of charged fermions~\cite{Georgi:1979df}. It is entirely possible to bypass this requirement by either judicious introduction of extra vector-like fermions~\cite{Witten:1979nr} or use of higher dimensional operators to correct charged fermion mass relations. However, both approaches have no unique implementation and, in the latter
case, the same class of operators could have significant effect on the gauge coupling unification~\cite{ Hill:1983xh,Shafi:1983gz}. To avoid these ambiguities we opt for the framework where viable charged fermion
masses are generated through 5- and 45-dimensional scalar representations at the tree-level and neglect influence of all higher dimensional operators. We accordingly discuss scalar fields in question in their most
natural setting -- a renormalizable SU(5) framework without SUSY -- in what follows.

The couplings of the multiplets in the $45$-dimensional scalar representation of $SU(5)$---$\bm{45}\equiv(\Delta_1, \Delta_2, \Delta_3, \Delta_4, \Delta_5,
\Delta_6, \Delta_7) = (\bm{8},\bm{2},1/2)\oplus
(\overline{\bm{6}},\bm{1}, -1/3) \oplus (\bm{3},\bm{3},-1/3)
\oplus (\overline{\bm{3}}, \bm{2}, -7/6) \oplus (\bm{3},\bm{1},
-1/3) \oplus (\overline{\bm{3}}, \bm{1}, 4/3) \oplus (\bm{1},
\bm{2}, 1/2)$---to the matter are set by a following pair of $SU(5)$ contractions: $(Y_1)_{ij}
(\bm{10}^{\alpha \beta})_i (\overline{\bm{5}}_{\delta})_j
\bm{45}^{*\,\delta}_{\alpha \beta}$ and  $\epsilon_{\alpha \beta \gamma \delta
\epsilon} (Y_2)_{ij} (\bm{10}^{\alpha \beta})_i (\bm{10}^{\zeta
\gamma})_j (\bm{4  5})^{\delta \epsilon}_\zeta$. $Y_1$ and $Y_2$ are Yukawa coupling matrices while the matter fields of the SM reside in $\bm{10}_i$ and $\overline{\bm{5}}_j$~\cite{Georgi:1974sy}, where $i,j=1,2,3$ are family indices. The couplings of $(\overline{\bm{3}},\bm{1}, 4/3)=\Delta_6$ are then 
\begin{eqnarray}
\mathcal L &\ni & (Y_1)_{ij} e_{i}^{C\,T} C  d^C_{a\,j}\Delta^{*}_{6\,a} \nonumber\\
&&+\sqrt 2 [(Y_2)_{ij} -(Y_2)_{ji}] \epsilon_{abc} u^{C\,T}_{i\,a} C u^C_{b\,j} \Delta_{6\,c}\,,
\end{eqnarray}
where $a,b=1,2,3$ are color indices. 
Note that the latter set of couplings is antisymmetric in flavour space. It is that fact that makes 
$\Delta_6$ leptoquarks innocuous as far as the proton decay is concerned at the tree level as has been 
noticed only recently~\cite{Dorsner:2009cu}.
Nevertheless, $\Delta_6$ could still mix via Higgs doublet 
with a scalar that has the right di-quark couplings to destabilize the proton~\cite{Weinberg:1980bf}. 
Indeed, there is one such scalar multiplet in the adjoint representation of 
$SU(5)$. However, in simple scenarios we have in mind, where there is only one adjoint scalar---$\bm{24}\equiv(\Sigma_8
, \Sigma_3, \Sigma_{(3,2)}, \Sigma_{(\overline{3},2)},
\Sigma_{24}) = (\bm{8},\bm{1},0)\oplus(\bm{1},\bm{3},0)
\oplus(\bm{3},\bm{2},-5/6)\oplus(\overline{\bm{3}},\bm{2},5/6)
\oplus(\bm{1},\bm{1},0)$---that breaks $SU(5)$, 
these particular components, i.e., $\Sigma_{(3,2)}$ and $\Sigma_{(\overline{3},2)}$, always get eaten by the so-called $X$ and $Y$ gauge bosons and are thus prohibited from mixing.

The couplings of $(\bm{8},\bm{2}, 1/2)=\Delta_1$ to the matter are 
\begin{eqnarray}
\mathcal L &\ni &-\sqrt{2} (Y_1)_{ij} d^T_{a\,i} (T^A)_{ab} C d^C_{b\,j} \Delta^{0\,A*}_{1}\,,\nonumber\\
&&-2 [(Y_2)_{ij} - (Y_2)_{ji}] u^T_{a\,i} (T^A)_{ab} C u^C_{b\,j}\Delta^{\,0A}_{1}\,,\nonumber\\
&&-\sqrt{2} (Y_1)_{ij} u^T_{a\,i} (T^A)_{ab} C d^C_{b\,j} \Delta^{+\,A*}_{1}\nonumber\\ 
&&+2 [(Y_2)_{ij} - (Y_2)_{ji}] d^T_{a\,i} (T^A)_{ab} C u^C_{b\,j} \Delta^{+\,A}_{1}\,.
\end{eqnarray} 
Here, $(T^A)_{ab}=1/2 (\lambda^A)_{ab}$, where $\lambda^A$ ($A=1,..,8$) are the usual Gell-Mann matrices of $SU(3)$. $\Delta_1$ has two sets of color components. One is electrically neutral---$\Delta_1^{0\,A}$---and the other is electrically charged---$\Delta_1^{+\,A}$. 

Clearly, both $\Delta_1$ and $\Delta_6$ have the right couplings to influence
the asymmetry we are interested in. However, in order to be relevant both $\Delta_1$ and $\Delta_6$ must be sufficiently light. This, on the other hand, has repercussions for unification of gauge couplings. In particular, $\Delta_6$ tends to either lower unification scale below proton decay limits or ruin gauge coupling unification all together if light enough. On the other hand, the $\Delta_1$ mass scale and the GUT scale, i.e., scale where the SM couplings unify, are inversely proportional. So,
whenever $\Delta_6$ is light $\Delta_1$ will also be light in order to keep the GUT scale above the limits imposed by proton decay. In other words, whenever $\Delta_6$ is light enough to play a role in low energy physics the same is true for $\Delta_1$.

We find that the simplest $SU(5)$ scenario comprising only $5$-, $24$- and $45$-dimensional scalar representations cannot unify at all with 
$\Delta_6$ light. (Admittedly, scenario with this content when $\Delta_6$ is sufficiently heavy unifies~\cite{Giveon:1991zm,Dorsner:2006dj,Dorsner:2007fy} but is already ruled out by proton decay experiments according to recent study~\cite{Dorsner:2007fy} unless one suppresses relevant operators due to either scalar~\cite{Dorsner:2009cu} or gauge boson exchange~\cite{Dorsner:2004xa} or both.) Of course, one can always judiciously add arbitrary number of additional representations of various dimensions to
adjust for that but we seek theoretically well-motivated scenario. We have accordingly checked unification with light $\Delta_6$ in the following two simple $SU(5)$ scenarios where all the Standard Model fermions have viable masses and mixing parameters. 

First, we have studied a scenario where, in addition to the $5$-, $24$- and $45$-dimensional scalar representations, 
we have a $15$-dimensional scalar representation~\cite{Dorsner:2006dj,Dorsner:2007fy} to generate neutrino masses via type II seesaw mechanism~\cite{Magg:1980ut,Schechter:1980gr,Lazarides:1980nt,Mohapatra:1980yp}. In this case, the gauge couplings of the SM do unify but the GUT scale comes out too low. Namely, upper bound turns out to be around $10^{13}$\,GeV which implies that the masses of $X$ and $Y$ gauge bosons---mediators of proton decay---are unacceptably small. 

We have then replaced the $15$-dimensional scalar representation with a
$24$-dimensional fermionic representation---$\bm{24}_F \equiv (\rho_8,\rho_3, \rho_{(3,2)}, \rho_{(\bar{3}, 2)},
\rho_{24})=(\bm{8},\bm{1},0)\oplus(\bm{1},\bm{3},0)\oplus(\bm{3},\bm{2},-5/6)
\oplus(\overline{\bm{3}},\bm{2},5/6)\oplus(\bm{1},\bm{1},0)$---to have renormalizable $SU(5)$ scenario~\cite{Perez:2007rm} where neutrino masses are generated via combination of type I~\cite{Minkowski:1977sc,Yanagida:1979as,GellMann:1980vs,Glashow:1979nm,Mohapatra:1979ia} and type III~\cite{Foot:1988aq,Ma:1998dn} seesaw mechanisms. (This is a renormalizable version of the model proposed in \cite{Bajc:2006ia} and further analyzed in \cite{Dorsner:2006fx}.) In that particular scenario situation is much more promising. Namely, the upper bound on the GUT scale comes out very close to the present experimental limits due to partial proton decay lifetime measurements with both $\Delta_1$ and $\Delta_6$ in the range accessible in collider experiments. Since the unification in this regime is very constraining and rich in features we discuss it in some details next.
\subsection{Unification}
For unification of the Standard Model (SM) gauge couplings at the GUT scale ($M_{GUT}$) to be successful  
at the one loop level two relevant equations~\cite{Giveon:1991zm} need to be satisfied:
\begin{subequations}
\begin{eqnarray}
\label{condition1} \frac{B_{23}}{B_{12}}&=&\frac{5}{8}
\frac{\sin^2
\theta_W-\alpha/\alpha_3}{3/8-\sin^2 \theta_W}=0.716 \pm 0.005,\\
\label{condition2} B_{12}&=&\frac{16 \pi}{5
\alpha} (3/8-\sin^2 \theta_W)=184.9 \pm
0.2.
\end{eqnarray}
\end{subequations}
The right-hand sides reflect the latest experimental
measurements of the SM parameters~\cite{Amsler:2008zzb} at $M_Z$ in the $\overline{MS}$
scheme: $\alpha_3 = 0.1176 \pm 0.0020$, $\alpha^{-1} = 127.906 \pm
0.019$ and $\sin^2 \theta_W = 0.23122 \pm 0.00015$.
The left-hand sides, on the other hand, depend on particle content and associated mass spectrum of the particular unification scenario. In fact, $B_{ij}=B_i - B_j$, where $B_i = \sum_{I} b_{iI} \ln
M_{GUT}/m_{I}$, $(M_Z \leq m_{I} \leq M_{GUT})$.
The sum over $I$ goes over all particles from $M_Z$---the SM particle scale---to the GUT scale. $b_{iI}$ are the usual $\beta$-function coefficients of the
particle $I$ of mass $m_I$. For example, the SM content with one light Higgs
doublet field has $b_1=41/10$, $b_2=-19/6$ and $b_3=-7$, and, accordingly, yields $B_{23}/B_{12}\approx0.53$.

Any potentially realistic $SU(5)$ grand unified
scenario must also allow for large enough masses of the $X$ and $Y$ gauge bosons since these vector leptoquarks mediate proton decay. Their practically degenerate masses are identified with $M_{GUT}$ ($m_{(X,Y)}\equiv M_{GUT}$), which in turn allows one to set lower bound on the GUT scale using proton decay lifetime limits. The most stringent bound comes from the latest experimental limit~\cite{:2009gd} on 
$p \rightarrow \pi^0 e^+$ partial decay lifetime: $\tau_{p \rightarrow \pi^0 e^+}> 8.2 \times 10^{33}$\,years. Theoretical prediction for this particular channel reads
\begin{equation}
\label{gamma}
\Gamma \approx \frac{m_p}{f^2_{\pi}} \frac{\pi}{4} A_L^2 
|\alpha|^2  (1 + D + F)^2  \frac{\alpha_{GUT}^2}{m^4_{(X,Y)}} \left[A^2_{S\,R} + 4 A^2_{S\,L} \right],
\end{equation}
where $A_{S\,L\,(R)}$ give a 
leading-log renormalization of the relevant operators
from the GUT scale to $M_Z$. These are given as~\cite{Buras:1977yy,Ellis:1979hy,Wilczek:1979hc}
\begin{equation}
A_{S\,L(R)}=\hspace{-0.2cm}\prod_{i=1,2,3} \hspace{-0.2cm} \prod_I^{M_Z \leq m_I \leq M_{GUT}}
\left[\frac{\alpha_i(m_{I+1})}{\alpha_i(m_I)}\right]^{\frac{\gamma_{L(R)i}}{\sum_J^{M_Z
\leq m_J \leq m_I} b_{iJ}}},
\end{equation}
where $\gamma_{L(R)i}=(23(11)/20,9/4,2)$.
The QCD running below $M_Z$ is captured by the
coefficient $A_L$. To establish lower bound on $M_{GUT}$ we further use $m_p=938.3$\,MeV, $D=0.81$, $F=0.44$,
$f_{\pi}=139$\,MeV, $A_L=1.25$ and
$|\alpha|=0.01$\,GeV$^3$~\cite{Aoki:2006ib}.

Also, the masses of all scalar leptoquarks that mediate proton decay must be sufficiently heavy to render the scenario viable. These are $\Psi_T$, $\Delta_3$ and  $\Delta_5$. (Note that $\Delta_6$ has erroneously been associated with a generation of the so-called $d=6$ proton decay operators~\cite{Weinberg:1980bf}.) If all the couplings of $\bm{45}$ with matter are taken into account~\cite{Dorsner:2009cu} and no suppression via Yukawa couplings is arranged their
masses should not be below $10^{12}$\,GeV due to the proton decay constraints. To remind us of that we place a line over
them in Table~\ref{tab:table1} for convenience where we list all nontrivial $b_i-b_j$ contributions of the scenario we analyze. 

\begin{widetext}

\begin{table}[t]
\caption{\label{tab:table1} The $b_i-b_j$ coefficient
contributions.}
\begin{ruledtabular}
\begin{tabular}{lcccccccccccccccc}
 & $\Psi_D$  & $\overline{\Psi_T}$ & $\Sigma_8$
 & $\Sigma_3$ & $\Delta_1$ & $\Delta_2$
 & $\overline{\Delta_3}$ & $\Delta_4$ & $\overline{\Delta_5}$
 & $\Delta_6$ & $\Delta_7$ & $\rho_8$
 & $\rho_3$ & $\rho_{(3,2)} + \rho_{(\bar{3}, 2)}$\\
\hline $b_{23}$ & $\frac{1}{6}$ & $-\frac{1}{6}$ &
$-\frac{3}{6}$ & $\frac{2}{6}$ & $-\frac{4}{6}$ & $-\frac{5}{6}$ &
$\frac{9}{6}$ & $\frac{1}{6}$ & $-\frac{1}{6}$ & $-\frac{1}{6}$ &
$\frac{1}{6}$ & $-\frac{12}{6}$ & $\frac{8}{6}$ & $\frac{4}{6}$\\
$b_{12}$ &$-\frac{1}{15}$ & $\frac{1}{15}$ & 0 &
$-\frac{5}{15}$ & $-\frac{8}{15}$ &$\frac{2}{15}$ &
$-\frac{27}{15}$ & $\frac{17}{15}$ & $\frac{1}{15}$
& $\frac{16}{15}$ &  $-\frac{1}{15}$ & $0$ & $-\frac{20}{15}$ & $\frac{20}{15}$\\
\end{tabular}
\end{ruledtabular}
\end{table}

\end{widetext}

In addition to these considerations there are two more constraint on the mass spectrum of the particles listed in Table~\ref{tab:table1}. Firstly, $\rho_8$ must be heavier than $10^{6}$\,GeV to accommodate the Big Bang Nucleosynthesis constraints~\cite{Bajc:2006ia}. 
Secondly, in the renormalizable model we are interested in~\cite{Perez:2007rm} there is a particular relation between masses of the fields in fermionic adjoint. Namely, due to small number of terms in the relevant part of the Lagrange density the masses of these fields are not independent from each other~\cite{Perez:2007rm}. As it turns out, we get unification only in one regime when $\Delta_6$ is light which corresponds to the following set of mass relations~\cite{Perez:2007rm}:
\begin{equation}
\label{condition4}
m_{\rho_8}=\hat{m}m_{\rho_3}, \qquad m_{\rho_{(3,2)}}=m_{\rho_{(\bar{3}, 2)}}=\frac{(1+\hat{m})}{2}m_{\rho_3},
\end{equation}
where $\hat{m}$ is a free parameter that basically represents a measure of the mass splitting between $\rho_8$ and $\rho_3$

With all these constraints in mind we are now in position to determine an upper bound on the
GUT scale at the one loop level assuming that $\Delta_6$ is responsible for asymmetry and is accordingly in the following mass range: $300$\,GeV~$\leq m_{\Delta_6} \leq 1$\,TeV (see section III for details). We do that by numerically maximizing $M_{GUT}$ while imposing that solution not only satisfies Eqs.~\eqref{condition1}, ~\eqref{condition2} and~\eqref{condition4} but that the masses given by numerical solution are in the following ranges: $10^2$\,GeV~$\leq m_{\Sigma_3}, m_{\Sigma_8}, m_{\Delta_1}, m_{\Delta_2}, m_{\Delta_4}, m_{\Delta_7}, m_{\rho_{(3,2)}}, m_{\rho_{(\bar{3}, 2)}} \leq M_{GUT}$, $10^{12}$\,GeV~$\leq m_{\Psi_T}, m_{\Delta_3}, m_{\Delta_5} \leq M_{GUT}$ and $10^6$\,GeV$\leq m_{\rho_8} \leq M_{GUT}$.

\begin{figure}
  \includegraphics[height=.24\textheight]{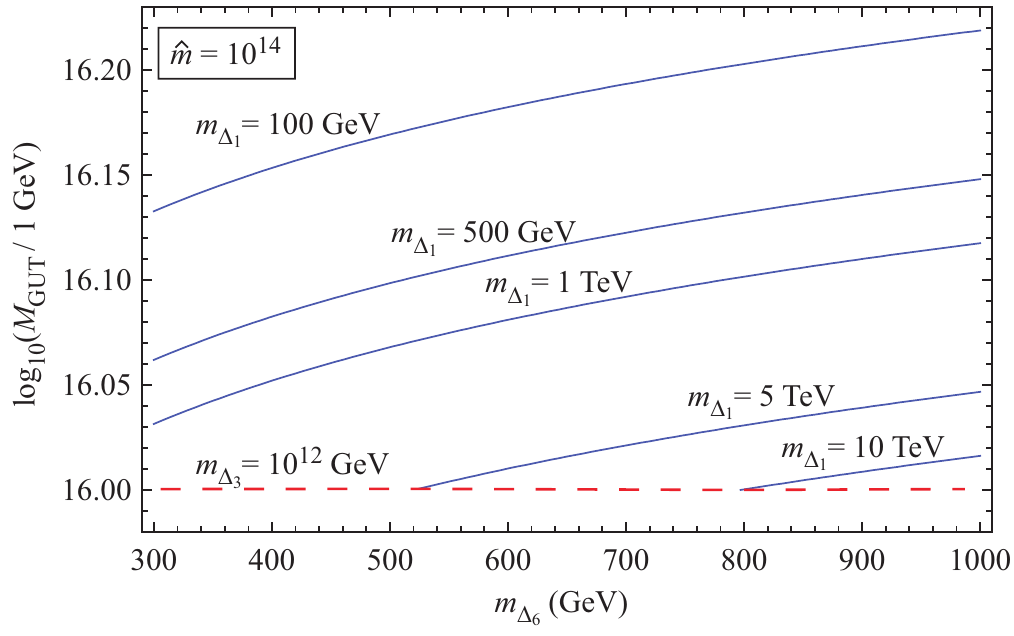}
  \caption{\label{figure:one} Viable unification in the $\hat{m} = 10^{14}$ case. Allowed region is between $m_{\Delta_1}=10^{2}$\,GeV and dashed line which is due to the constraint imposed by experimental results on proton decay when imposed on the $p$ decay predictions due to the scalar boson exchange.}
\end{figure}

\begin{figure}
  \includegraphics[height=.24\textheight]{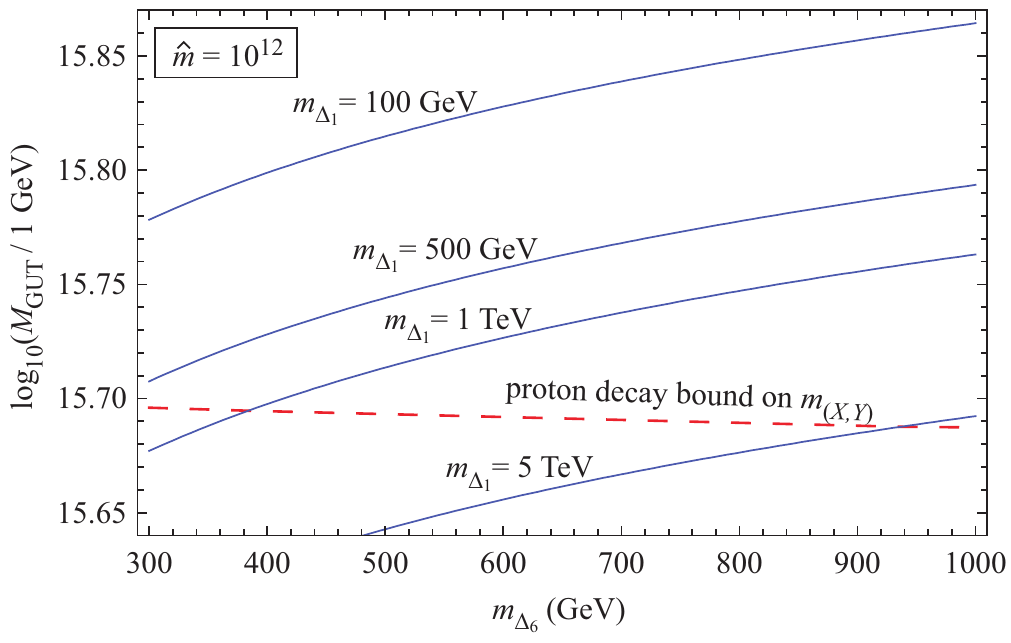}
  \caption{\label{figure:two} Viable unification in the $\hat{m} = 10^{12}$ case. Phenomenologically allowed region is between $m_{\Delta_1}=10^{2}$\,GeV and dashed line which stands for the lower bound on $M_{GUT}$ due to the $d=6$ proton decay via gauge boson exchange.}
\end{figure}

In figure~\ref{figure:one} (figure~\ref{figure:two}) we present the $\hat{m}=10^{14}$ ($\hat{m}=10^{12}$) scenario. 
We opt for this because in the region of parameter space we are interested in, i.e., $300$\,GeV$\leq m_{\Delta_6} \leq 1$\,TeV, viable unification can exist only when $10^{6.4} \leq \hat{m} \leq 10^{14.2}$. However, if $M_{GUT}$ is to be above the limit imposed by proton decay measurements then $10^{11.5} \leq \hat{m} \leq 10^{14.2}$.

It is important to stress that for each viable unification point in figures~\ref{figure:one} and~\ref{figure:two} we know the full mass spectrum of the model including $\alpha_{GUT}$ and $A_{S\,L(R)}$. We can thus establish an accurate lower bound on $M_{GUT}$ by matching the prediction of the scenario for partial lifetime of $p \rightarrow \pi^0 e^+$ at the leading order due to the gauge boson exchange with the current experimental limit. This prediction is seen in figure~\ref{figure:two} as a dashed line. Space below that line should be considered as excluded. In figure~\ref{figure:one}, however, an almost horizontal dashed line represents a line along which the mass of proton decay mediating scalar $\Delta_3$ is at its experimentally imposed limit ($m_{\Delta_3}=10^{12}$\,GeV). Again, space below that line should be considered as excluded. What happens is that as $\hat{m}$ is varied we go from one regime where the so-called $d=6$ proton decay operators due to gauge boson exchange dominate ($\hat{m}=10^{14}$) to the other regime where the $d=6$ proton decay operators due to scalar exchange dominate. The important consequence of this flip in proton decay dominance is that the entire viable region of unification when $300$\,GeV$\leq m_{\Delta_6} \leq 1$\,TeV will be experimentally excluded with an improvement of about a factor of 6 in partial lifetime measurement for $p \rightarrow \pi^0 e^+$. This is what makes regime of light $\Delta_6$ extraordinarily predictive and testable.

Clearly, for a given value of $m_{\Delta_6}$ (and $\hat{m}$) we have an upper bound on $m_{\Delta_1}$. For example, when $m_{\Delta_6}=520$\,GeV and $\hat{m}=10^{14}$ (figure~\ref{figure:one}) we have $m_{\Delta_1} \leq 5$\,TeV. 

\begin{figure}
  \includegraphics[height=.24\textheight]{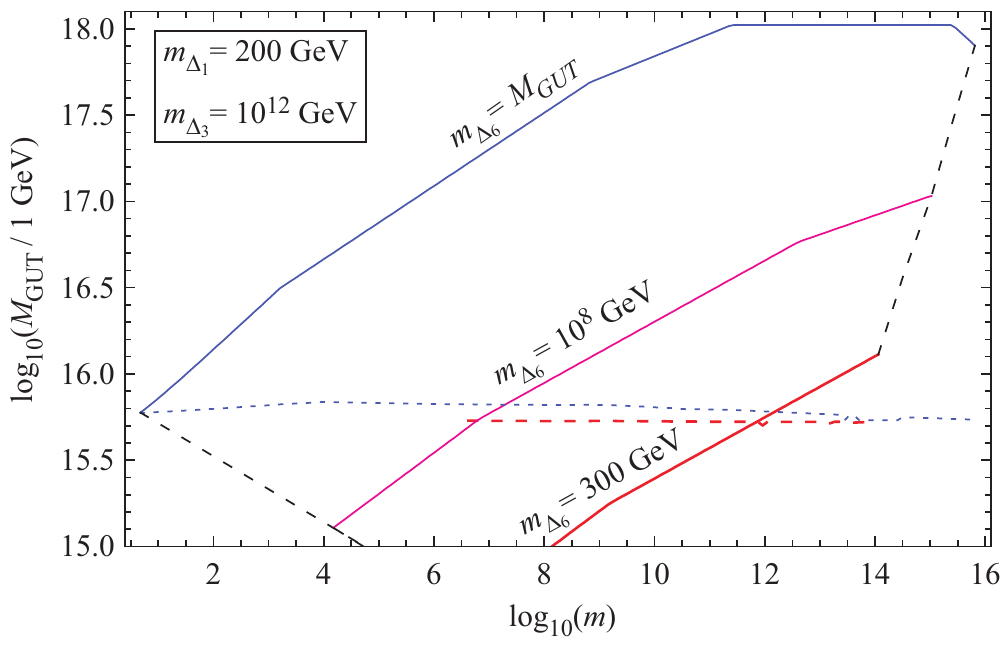}
  \caption{\label{figure:three} Upper bound on $M_{GUT}$ as a function of $\hat{m}$ for different values of $m_{\Delta_6}$.}
\end{figure}

Our findings on unification in the setup with $5$-, $24$- and $45$-dimensional scalar representations and one adjoint fermionic representation substantially differ from what has been presented elsewhere~\cite{Perez:2008ry}. In particular, if $\Delta_6$ is not assumed to be light we find no reasonably accessible upper bound on $M_{GUT}$ nor any upper bound on $m_{\Delta_1}$, contrary to what has been put forth in ref.~\cite{Perez:2008ry}. To illustrate our disagreement with respect to the $M_{GUT}$ predictions we plot in figure~\ref{figure:three} what we find to be upper bound on the GUT scale as a function of $\hat{m}$ for varying values of $m_{\Delta_6}$. For definiteness we want to compare our finding with those of ref.~\cite{Perez:2008ry} when the masses in the fermionic adjoint obey eq.~\eqref{condition4} and $m_{\Delta_1}=200$\,GeV, $m_{\Delta_3}=10^{12}$\,GeV and $m_{\Delta_6}=M_{GUT}$. Clearly, upper bound on $M_{GUT}$ we find when $m_{\Delta_6}=M_{GUT}$ depends strongly on $\hat{m}$
and substantially exceeds the value of $7.8 \times 10^{15}$\,GeV that has been advertised in ref.~\cite{Perez:2008ry} as an absolute upper bound in that case. Moreover, the corresponding experimental bound due to proton decay on $M_{GUT}$ (dotted line) for the $m_{\Delta_6}=M_{GUT}$ case is more than two orders of magnitude below predicted value for $M_{GUT}$ in substantial part of available parameter space. Only when $\hat{m}<1$ do we actually get scenario that can be probed by proton decay.

If, however, we assume that $m_{\Delta_6}$ is light---as we do---the picture changes completely. Red solid line represents upper bound on $M_{GUT}$ for $m_{\Delta_6}=300$\,GeV while the dashed line represents corresponding proton decay limit on the GUT scale. Clearly, significant parts of parameter space in this regime are already excluded by proton decay experiments. (Note also that the unification is not possible for all values of $\hat{m}$. Both upper and lower bounds on $\hat{m}$ as a function of $m_{\Delta_6}$ are given as thin dashed lines.)

We trace the above mentioned disagreement to a simple fact that the analysis in ref.~\cite{Perez:2008ry} does not vary all the masses of the fields in the model within their phenomenologically allowed ranges but makes judicious choice of varying only those masses that are associated with fields with negative $b_{12}$ coefficients and members of the adjoint fermionic representation. What we have presented is a consistent unification analysis at the one loop level. If one is to consider two loop effects in the running of gauge couplings the relevant range of allowed masses for $m_{\Delta_6}$ and $m_{\Delta_1}$ would slightly change but they would still be within the TeV range.



\section{The $t\bar t$ production cross section and the forward-backward asymmetry}

In this section we investigate the impact of light $\Delta_{1,6}$ states on the production of $t\bar t$ pairs at the Tevatron. The analysis is performed in a model independent way and the results apply to any model scenarios with light colored scalar triplets and/or octets. The current versions of widely used Monte Carlo tools for colliders such as MadGraph/MadEvent and CalcHEP cannot handle the color flow of the required $\Delta_6$ couplings \cite{Kang:2007ib}. Therefore we consider LO inclusive production cross sections on which we do not impose any kinematical cuts. A more realistic evaluation would also need to properly simulate particle decays, hadronic showering of their products and detector effects, all of which can in principle affect the quantities under study.  However, we have checked that our calculation of the $t\bar t$ production in the SM agrees well with results from MadEvent. This gives us confidence in the reliability of our crude approach.

\begin{figure}[th]
\begin{center}
\includegraphics[width=8.5cm]{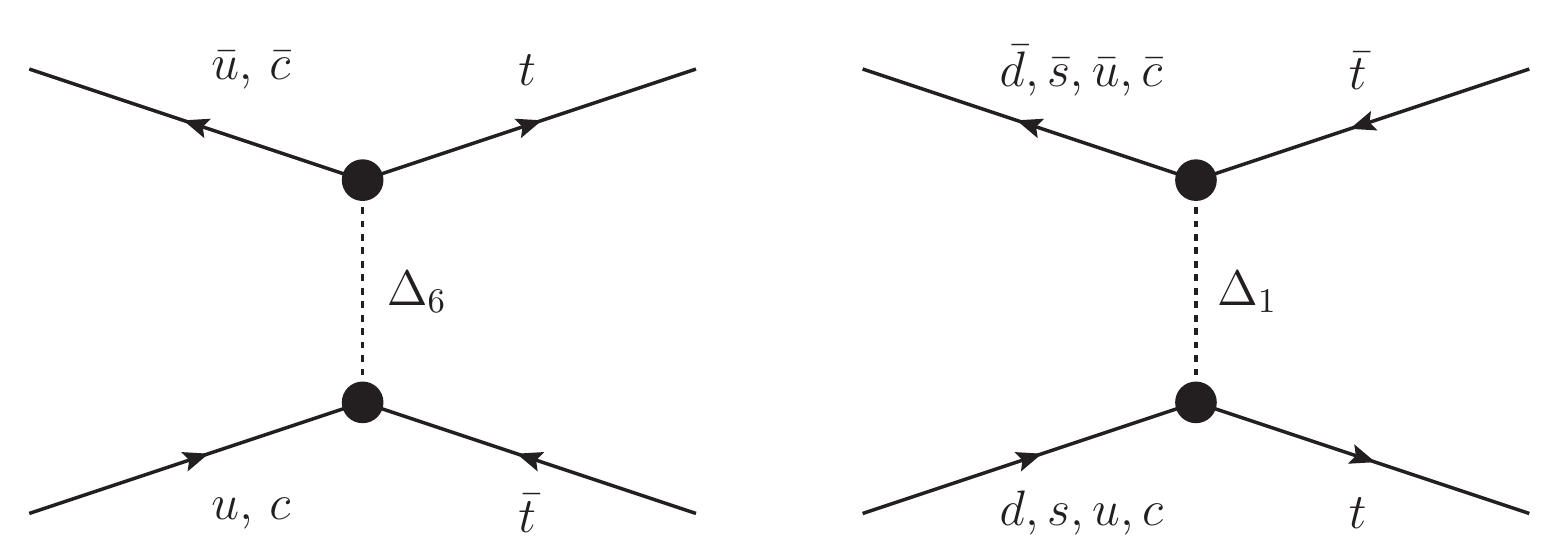}
\end{center}
\caption{ \label{fig:0b} Leading contributions to $t\bar t$ production cross section and the FBA at the Tevatron coming from $\Delta_6$ (left) and $\Delta_1$ (right) exchange. }
\end{figure}

The $\Delta_6$ contributes to the $t\bar t$ production from u-channel exchange in left diagram on figure \ref{fig:0b}. It interferes with the SM $q\bar q$ contribution for the $u\bar u$ and $c\bar c$ initial partons yielding
\begin{eqnarray}
\frac{d\sigma^{q\bar q}_6(\hat s)}{d\hat t}  &=& \frac{d\sigma_{SM}^{q\bar q}(\hat s)}{d \hat t} - \frac{  \alpha_s {|g^{qt}_6 |}^2}{9  \hat s^3}  \frac{\left(m_t^2 \hat s + (m_t^2-\hat u)^2\right)}{ \left(m_{\Delta_6} ^2-\hat u\right)}  \nonumber\\
&&+ \frac{{|g^{qt}_{6} |}^4 }{48 \pi  s^2 } \frac{ \left(m_t^2-\hat u\right)^2}{\left(m_{\Delta_6} ^2-\hat u\right)^2} \,,
\label{eq:sigmaLQ}
\end{eqnarray} 
where $\hat t=(p_{u}-p_t)^2$, $\hat u=(p_{\bar u}-p_t)^2$ and we have denoted the relevant parameters of the $\Delta_6$ as $g_6^{qt}\equiv 2 \sqrt 2 (Y_2^{qt}-Y_2^{tq})$ for $q=u,c$. 
Similarly the $\Delta_1$ contribution can be obtained by reversing the color and fermion flow of t quarks as seen in the right diagram of figure \ref{fig:0b} and adjusting the required color factors. The differential cross section can then be written as
\begin{eqnarray}
\frac{d\sigma^{q\bar q}_1(\hat s)}{d\hat t}  &=& \frac{d\sigma_{SM}^{q\bar q}(\hat s)}{d \hat t} + \frac{2  \alpha_s {|g^{qt}_1 |}^2}{27  \hat s^3}  \frac{\left(m_t^2 \hat s + (m_t^2-\hat t)^2\right)}{ \left(m_{\Delta_1} ^2-\hat t\right)}  \nonumber\\
&&+  \frac{ {|g^{qt}_{1} |}^4 }{18 \pi  s^2 } \frac{ \left(m_t^2-\hat t\right)^2}{\left(m_{\Delta_1} ^2-\hat t\right)^2} \,,
\label{eq:sigmaLQ1}
\end{eqnarray} 
where now also the down quark partons contribute to the $\Delta_1$ mediated cross section terms. We denote  $g_1^{ut}\equiv 4 (Y_2^{ut}-Y_2^{tu})$ and $g_1^{dt}\equiv 4 \sqrt{  (Y_2^{dt}-Y_2^{td})^2 + Y_1^{dt*2}/ 8}$, where $u$ actually stands for $u,c$ quark flavours and $d$ for $d,s$.  In our analysis we neglect possible interference contributions between both $\Delta$'s. 

In order to obtain the hadronic cross section, we convolute the partonic result with the appropriate parton distribution functions (PDFs) and consistently transform the phase-space integration from the parton to the lab frame. We obtain
\begin{equation}
\frac{d\sigma(s)}{dt} = \hspace{-0.2cm}\sum_{p,p'=q,g} \int_{x_0}^1 dx_1 \int_{x_0}^1 dx_2  x_1 x_2 \frac{d\sigma^{pp'}(\hat s)}{d\hat t} f_p(x_1) f_{p'}(x_2)\,,
\end{equation}
where $f_p(x)$ is the (anti)proton PDF for parton $p$,  $x_0 = 4m_t^2 / s$ is the physical PDF threshold cutoff for our process, $\hat s = x_1 x_2 s$ and $\hat t = x_1 x_2 (t-m_t^2) +m_t^2$ is the transformed Mandelstam variable $\hat t=(p_t - p_{p'})^2$. The factor $x_1 x_2$ in the integrand is the corresponding Jacobian. In the $x_1,x_2$ integration we have to furthermore impose kinematical limits $\hat s > 4m_t^2$ and  $-\hat s (1 + \hat \beta_t)/2 + m_t^2 < \hat t < -\hat s (1 - \hat \beta_t)/2 + m_t^2 $, where $\hat \beta_t = \sqrt{1-4m_t^2/\hat s}$. 
The sum runs over all $q$ quark as well as anti-quark flavours and the gluon contribution, however as discussed above, relevant leading contributions only involve diagonal $p=\bar p'=q$ and (in the SM) $p=p'=g$ entries. Finally, to obtain the polar angle ($\theta$) distribution of $t \bar t $ pairs, we transform the above differential cross section to
\begin{equation}
\frac{d\sigma(s)}{d\cos \theta} = \frac{s \beta_t }{2}  \frac{d\sigma(s)}{dt(\cos\theta)}\,,
\end{equation}
where $t(\cos\theta) = -s (1-\cos \theta \beta_t)/2 + m_t^2$. The FBA and the total inclusive cross section are then defined as
\begin{eqnarray}
A^{t\bar t}_{FB}(s) &=& \frac{\int_0^1 d\cos\theta \left[\frac{d\sigma^{}(s)}{d \cos\theta}\right] - \int_{-1}^0 d\cos\theta \left[\frac{d\sigma^{}(s)}{d \cos\theta}\right]}{\sigma_{t\bar t}(s)} \,,\nonumber\\
&&\\
\sigma_{t\bar t}(s) &=& \int_{-1}^1 d\cos\theta \left[\frac{d\sigma^{}(s)}{d \cos\theta}\right]\,.
\end{eqnarray}

\begin{figure}[t]
\begin{center}
\includegraphics[width=8.5cm]{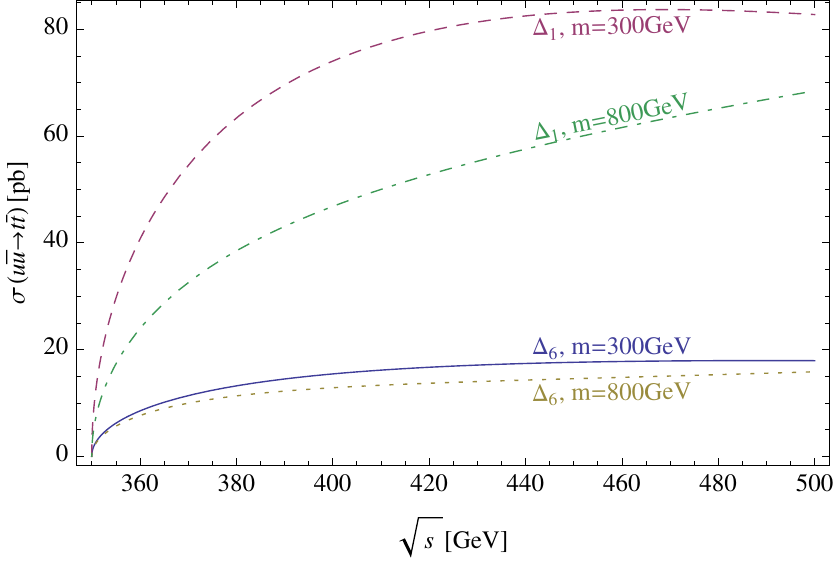}
\includegraphics[width=8.5cm]{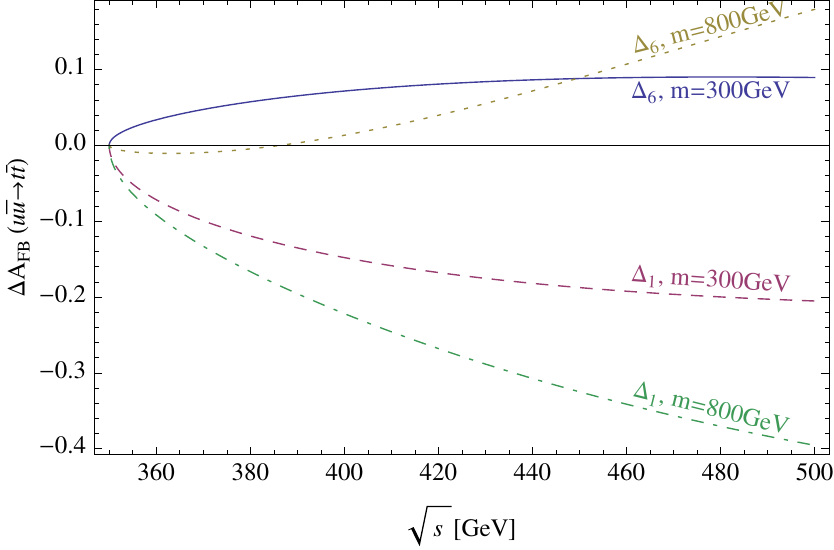}
\end{center}
\caption{ \label{fig:0.5}Examples of the partonic $t\bar t$ cross section (top) and the contribution to the FBA (bottom), induced by the exchange of $\Delta_{1,6}$.}
\end{figure}

Applying these formulae at the partonic level, it becomes clear that the couplings of the octet $\Delta_1$ cannot induce a large positive FBA. This is also seen on figure \ref{fig:0.5}, where we plot examples of the partonic cross section and the FBA induced by the two $\Delta$'s, for two different $\Delta_{1,6}$ masses. 
The $\Delta_1$ contributions interfere constructively with the SM amplitude resulting in big enhancements in the cross section, while the induced FBA is negative. On the other hand, $\Delta_6$ tends to produce a large positive asymmetry at large values of partonic center of mass energy, while the induced asymmetry is negative close to the threshold. Therefore, in order to obtain a good fit to both the cross section and the FBA the contributions of $\Delta_1$ should generically be suppressed compared to those of $\Delta_6$. This would be the case if $\Delta_1$ was much heavier than $\Delta_6$. However, this possibility is limited in this model, by the existing proton decay bounds. Consequently, some fine-tuning is needed between the couplings of $\Delta_1$ to up and down quarks in order to suppress its effects.  

In our numerical analysis we use the CTEQ5 \cite{Lai:1999wy} set of PDF's at the single renormalization and factorization scale $\mu_F=\mu_R=m_t$ at which we also evaluate the strong coupling constant $\alpha_s(m_t) = 0.108$. We use $m_t = 175$ GeV -- the value used by the CDF analysis \cite{CDFNote9448}, and rescale our results so that our (tree-level) SM value agrees with the SM prediction at NLO in QCD for $\sigma_{t\bar t}(s)$ \cite{ttbarXsecTheo}. We apply the same procedure for each bin when looking at the invariant $t\bar t$ mass ($m_{t\bar t}$) spectrum and take the reference SM predictions from \cite{Frederix:2007gi}. 

\begin{figure}[t]
\begin{center}
\includegraphics[width=8.5cm]{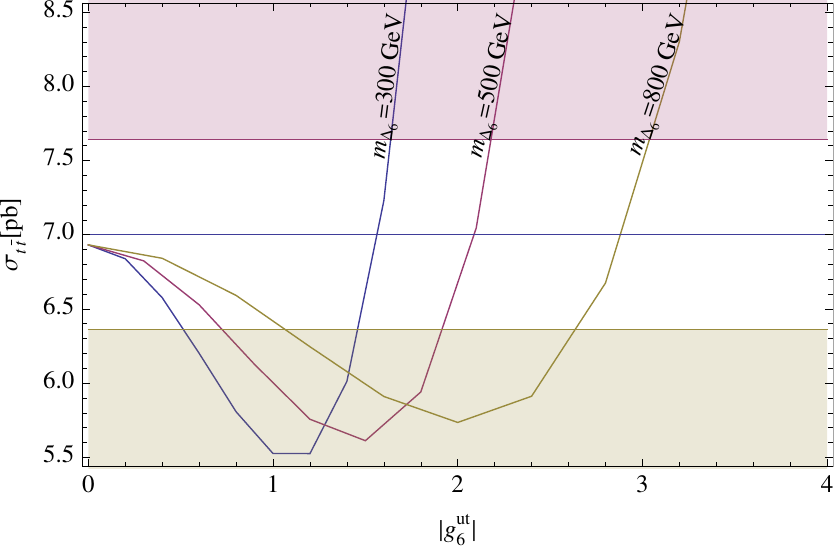}\quad
\includegraphics[width=8.5cm]{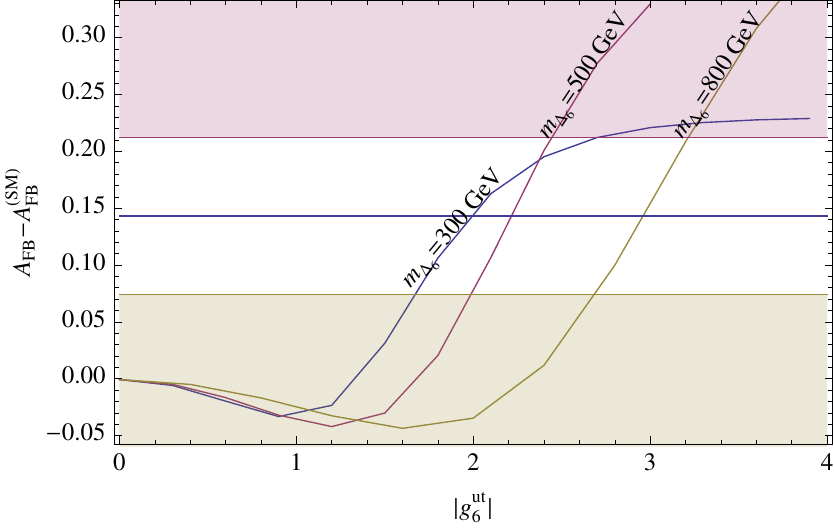}
\end{center}
\caption{ \label{fig:1} Examples of the hadronic $t\bar t$ cross section and the FBA at the Tevatron including $\Delta_6$ contributions. The shaded regions are outside the one sigma experimental bounds. For the FBA, the SM contribution is subtracted from the plotted values.}
\end{figure}

\begin{figure}[th]
\begin{center}
\includegraphics[width=8.5cm]{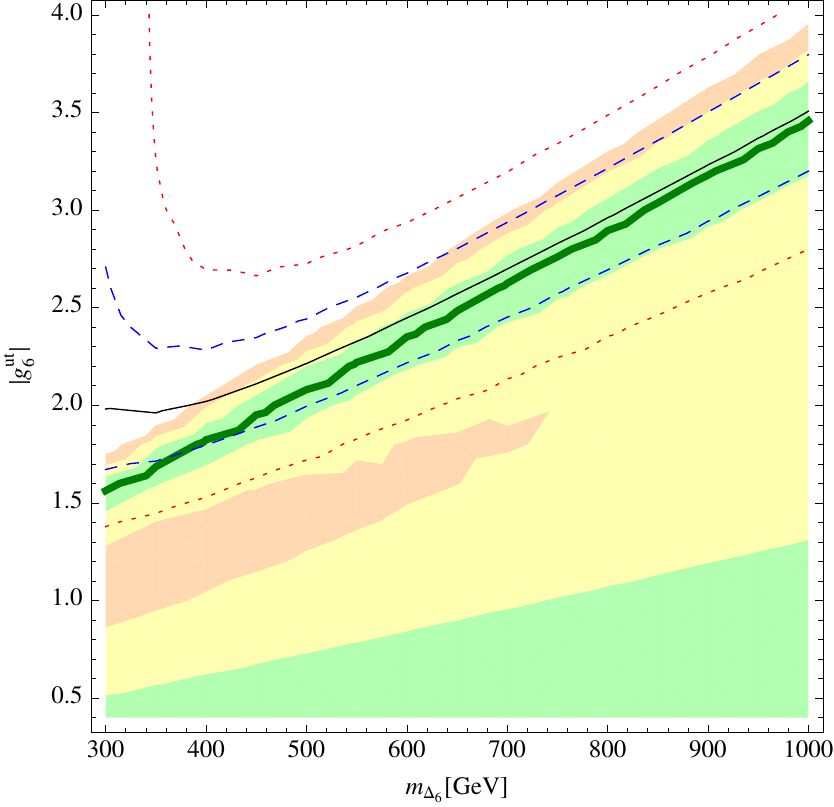}
\end{center}
\caption{ \label{fig:2} Combined constraints on the parameter space of $\Delta_6$ given by the  integrated $t\bar t$ production cross section (shaded regions) and the total FBA (contours). The $68,95,99\%$ confidence level regions in production cross section are shaded in green, yellow and orange respectively. The corresponding $68(95)\%$ CL regions in the FBA are bounded by blue dashed (red dotted) contours. The best-fit contours are drawn in thick (thin) full lines for the cross section and the FBA respectively.}
\end{figure}

We plot the cross section and the FBA at the Tevatron in figure \ref{fig:1} as a function of the $g_6^{ut}$ coupling for three $\Delta_6$ masses. We compare the FBA to the difference between the measured value and the SM prediction $A^{exp}_{FB}-A_{FB}^{SM}=  (14.2 \pm 6.9) \%$, while we use the CDF cross section value of $\sigma_{t\bar t}^{exp} = 7.0\pm 0.6$ pb \cite{CDFNote9448}.
We see that the values for the $g_6^{ut}$ coupling, required to explain the measured FBA are quite large and are positively correlated with $m_{\Delta_6}$ so that the mass of $\Delta_6$ should be right at the electroweak scale to avoid problems with pertubativity. 
While $\Delta_6$ also contributes in the $c\bar u$ and $u\bar c$ channels, there is no interference with the leading SM contribution, so that only the last term in eq. (\ref{eq:sigmaLQ}) would remain with the coupling replacement $|g_6^{qt}|^4 \to |g_6^{ct}g_6^{ut*}|^2$. Due to the small $c$ and $\bar c$ PDF components these contributions are further suppressed. For the same reason the $c\bar c$ component contribution cannot significantly affect the cross section for reasonable values of $\Delta_6$ parameters and we are left with the $u \bar u$ contribution.
The combined constraints on the parameter space of the model given by the integrated $t\bar t$ production cross section and the total FBA are shown in figure \ref{fig:2}.
We obtain that both the $t\bar t$ production cross section and the FBA can be accommodated simultaneously due to the negative interference contribution to the cross section, provided  $m_{\Delta_6} \geq 300$ GeV. 
The corresponding best-fit relation between the parameters $g_6^{ut}$ and $m_{\Delta_6}$ in this region can be put into the approximate form
\begin{equation}
|g_6^{ut}| = 0.9(2) + 2.5(4) \frac{m_{\Delta_6}}{1\,\mathrm{TeV}}\,.
\label{eq:fit}
\end{equation}

\begin{figure}[th]
\begin{center}
\includegraphics[width=8.5cm]{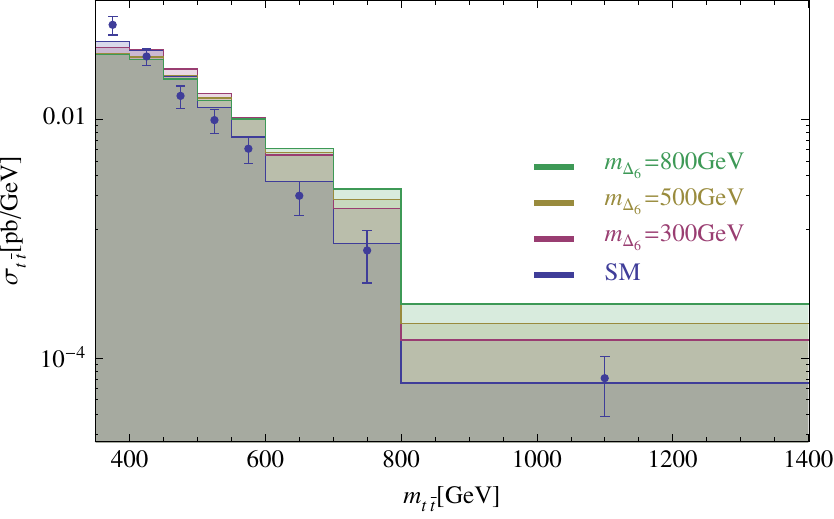}

\vspace{0.3cm}

\includegraphics[width=8.5cm]{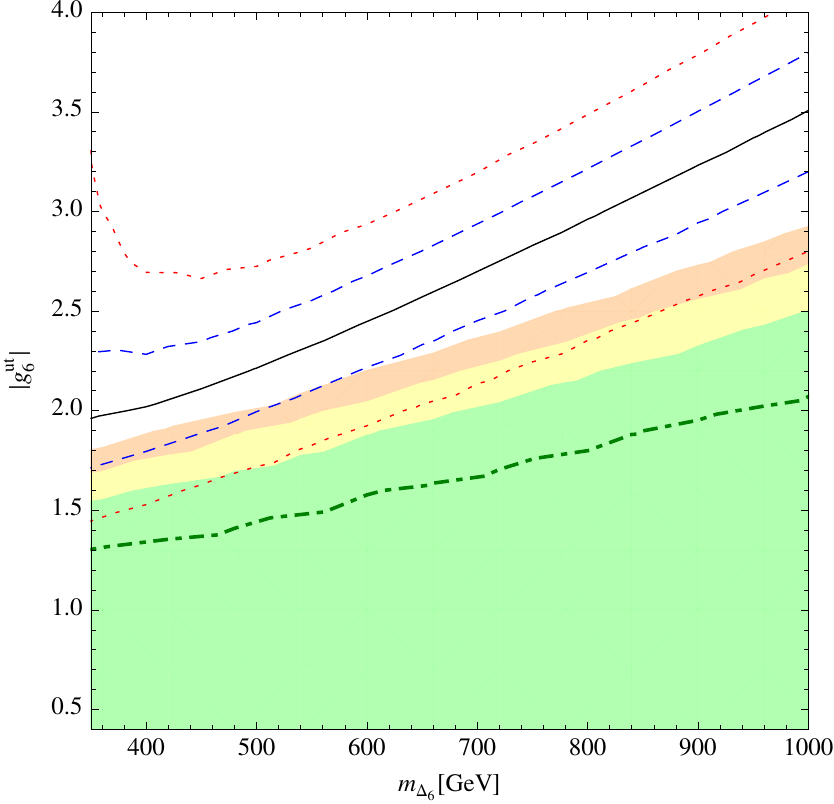}
\end{center}
\caption{ \label{fig:3} $\Delta_6$ contributions to the $m_{t\bar t}$ invariant mass spectrum in $t\bar t$ production at the Tevatron (top) and the resulting constraints on the parameter space (bottom). In the top plot, values of $g_{6}^{ut}$ are chosen according to eq. (\ref{eq:fit}).  In the bottom plot, the constraint on the parameter space of $\Delta_6$ is given by the highest invariant mass bin in $t\bar t$ production (shaded regions).  The $68, 95, 99\%$ confidence level regions are shaded in green, yellow and orange respectively. The best fit contour is plotted in green dot-dashed line. Superimposed are the $68(95)\%$ confidence region contours of the total FBA as in figure \ref{fig:2}.}
\end{figure}

Another important constraint on the relevant parameters of $\Delta_6$ in $t\bar t$ production, in particular its mass, comes from the high $m_{t\bar t}$ spectrum measured by the CDF collaboration \cite{Aaltonen:2009iz} which reports $\sigma^{t\bar t}(800 \,\mathrm{GeV} \leq m_{t\bar t}\leq 1400\, \mathrm{GeV}) = 0.041 \pm 0.021$ pb, where we have given the integrated cross section in the highest bin and combined the experimental errors of \cite{Aaltonen:2009iz} in quadrature.
This constraint is shown in figure \ref{fig:3} together with the contours of the FBA fit.
We see that there is some tension between this constraint and the FBA fit throughout the relevant mass range. The most interesting region lies around $m_{\Delta_6} \approx 400$ GeV where the tension is the smallest. It grows stronger for larger $\Delta_6$ masses and the two observables cannot both be reconciled for $m_{\Delta_6} $ above the TeV scale.  
However, due to the questionable reliability of our naively corrected LO estimation at high $m_{t\bar t }$, this bound is to be considered as tentative at the moment. A more decisive conclusion would require a consistent evaluation of the observable at NLO in QCD including the NP contributions which is beyond the scope of this analysis.

Note that a similar constraint can also be derived from the recent CDF measurement of the $m_{t\bar t}$ spectrum of the FBA \cite{CDFNote9853}, which exhibits large positive contributions { in the low top-pair invariant mass region}. This poses a problem for any NP explanation of the asymmetry with heavy mediators in the $t(u)$-channels, since it will generically predict a small FBA in this region of phase space. At the moment, the significance of such a constraint is still lower than the high end tail of the production cross section spectrum, however it is potentially much more robust, since it is believed to be less sensitive to higher order QCD effects and PDF uncertainties \cite{AFBtheo}. 


\section{Other constraints}

\begin{figure}[th]
\begin{center}
\includegraphics[width=8.cm]{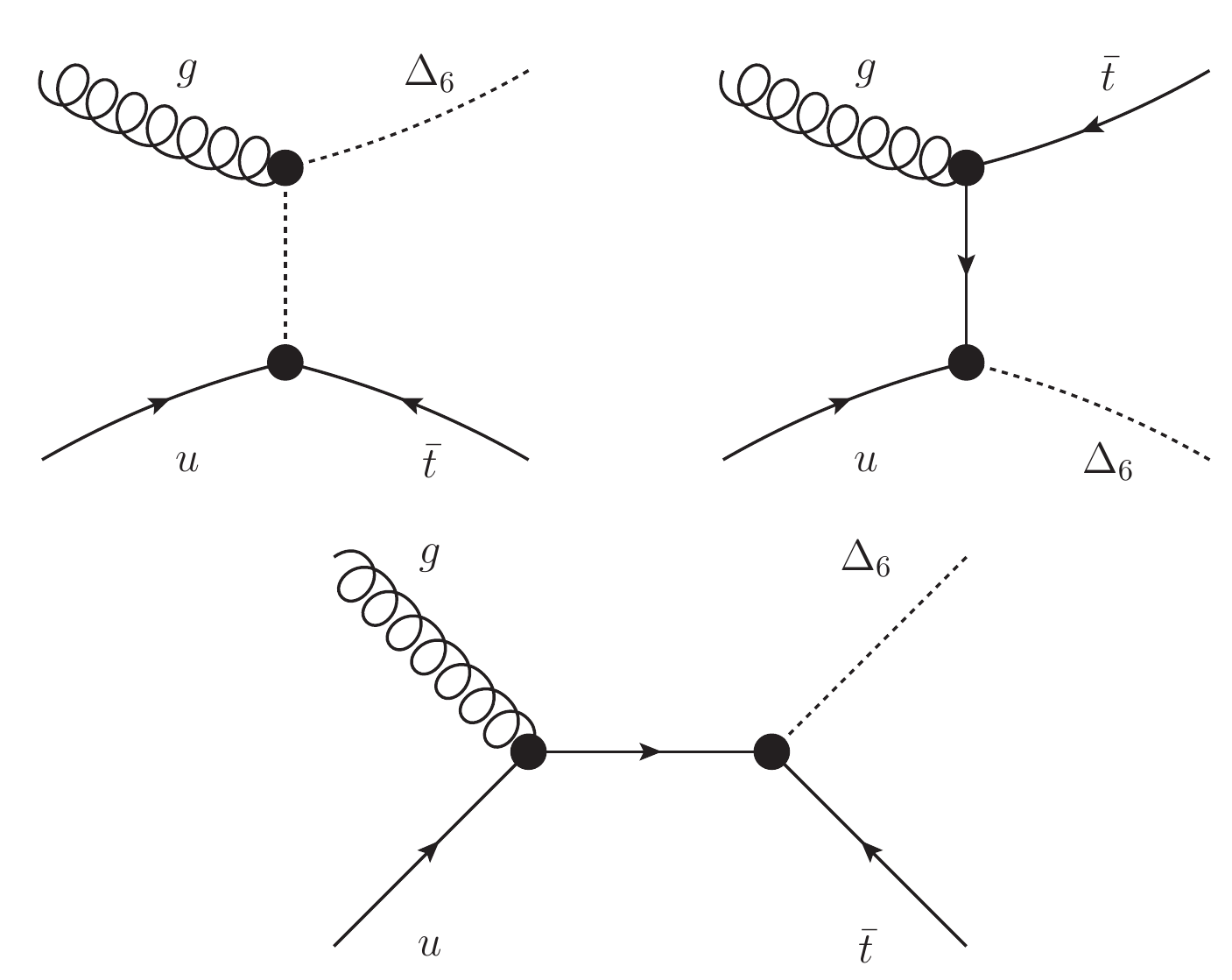}
\end{center}
\caption{ \label{fig:4} Diagrams contributing to partonic single $\Delta_6$ production, associated with an (anti)top quark.}
\end{figure}
\begin{figure}[th]
\begin{center}
\includegraphics[width=8.5cm]{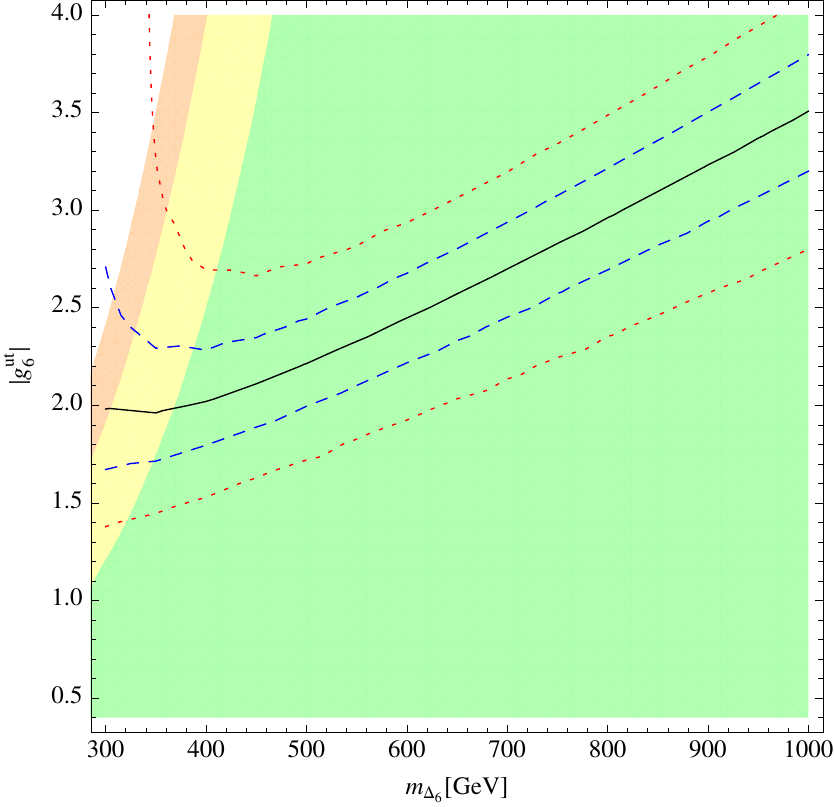}
\end{center}
\caption{ \label{fig:ttj} Constraint on the parameter space of $\Delta_6$ coming from $t\bar t$+jet production (shaded regions).  The $68, 95, 99\%$ confidence level regions are shaded in green,yellow and orange respectively. Superimposed are the $68(95)\%$ confidence region contours of the total $t\bar t$ production FBA as in figure \ref{fig:2}.}
\end{figure}

The constraints on color octets such as $\Delta_1$ have been considered elsewhere in some detail~\cite{octets} and we will not repeat the discussion here, noting only that they are allowed to be as light as $100$~GeV. Regarding $\Delta_6$, the most robust lower limit on its mass is given by the high energy run of LEP II, putting the lower bound at roughly $m_{\Delta_6}>105$ GeV~\cite{Arnold:2009ay}. Tevatron searches for resonances in the invariant mass spectrum of di-jets \cite{Aaltonen:2008dn} only constrain the $g_6^{cu}$ coupling. A sizable $g_6^{tu}$ coupling does not induce FCNC processes unless one of the other $g_6^{ij}$ couplings is different from zero as well. Even in their presence, contributions to FCNC processes of first-two generation up quarks such as $D - \bar D$ mixing only appear at one loop and we leave this for a subsequent study. 
Contrary to neutral t-channel mediators of $t\bar t$ production, the $\Delta_6$ does not induce like-sign top pair production which would otherwise severely constrain the high $\Delta_6$ mass region. In addition, the single top production cross section necessarily involves the $g_6^{uc}$ coupling which can be suppressed. On the other hand, an important constraint on the $g^{tu}_6$ comes from the production of $t\bar t $+jet which has been measured by CDF in agreement with the SM predictions \cite{CDFNote9850}. The $\Delta_6$ contributes to this process through its associated production with a top or an anti-top quark. The contributing diagrams for the latter are shown in figure \ref{fig:4}.
The contribution to $\sigma_{t\bar t + j}$ can then be written approximately as $\sigma^{\Delta_6}_{t\bar t + j} \approx (\sigma_{t\Delta^*_6} + \sigma_{\bar t \Delta_6}) \times Br(\Delta_6 \to t u)$. The expression for the underlying partonic cross section $\sigma({u g \to \bar t\Delta_6})$ 
is rather lengthy and is given in the appendix. In case we put all other couplings of $\Delta_6$ to zero, its branching ratio to $t u$ pairs will be one for $\Delta_6$ masses above the top. The above (narrow width) approximation is valid in the whole interesting parameter region provided no other channels contribute significantly to the $\Delta_6$ width. We are also neglecting possible interference effects with the SM. This is a reasonable approximation at the Tevatron, where the dominant SM contributions  come from gluon radiation and do not interfere with the $\Delta_6$ contribution. The resulting constraint on parameters of the model is shown in figure \ref{fig:ttj}.
We see that the constraint, although not competitive at the present level of experimental precision, is potentially important for low $\Delta_6$ masses.


\section{Search strategies}

A detailed analysis of $\Delta_1$ phenomenology at colliders has already been performed in refs. \cite{octets,FileviezPerez:2008ib}, where it was found that such states could be observed at the LHC for masses around or below TeV. As shown in the previous sections, this is also the case in our scenario, provided that the $\Delta_6$ is responsible for the observed deviation in the FBA.

\begin{figure}[th]
\begin{center}
\includegraphics[width=8.5cm]{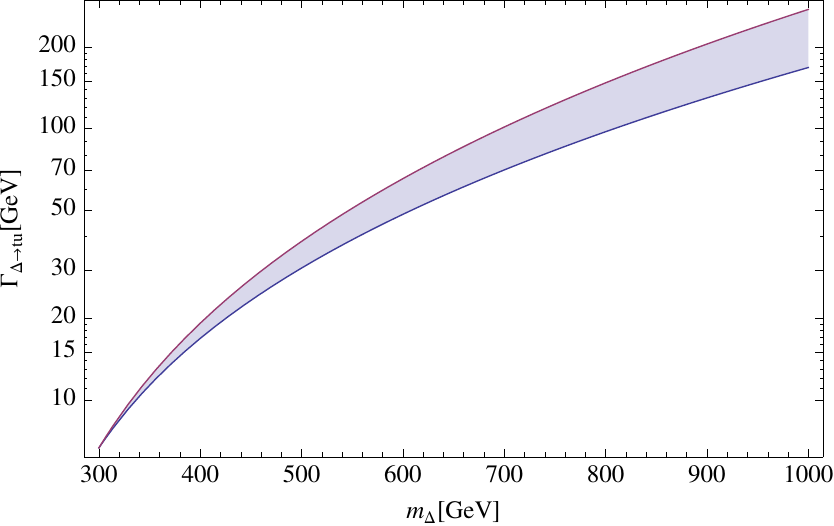}
\end{center}
\caption{ \label{fig:5} Dependence of the minimal $\Delta_6$ width on the mass of ${\Delta_6}$.  Values of $g_{6}^{ut}$ are chosen according to eq. (\ref{eq:fit}).}
\end{figure}

Regarding $\Delta_6$, in the interesting mass region its width is
dominated by the two body decay ${\Delta_6} \to t u$ which can be written as
 \begin{equation}
\Gamma(\Delta_6 \to t u) = \frac{|{g^{ut}_{6}}|^2 \left(m_{\Delta_6}^2- m_t^2\right)^2}{16 \pi  m_{\Delta_6}^3}\,.
 \end{equation}
If $\Delta_6$ has no other relevant interactions and we neglect loop-induced decay channels, the dependence of $\Delta_6$ width on its mass is shown in figure \ref{fig:5}, where we have again taken values of $g_6^{ut}$ which reproduce the FBA within one standard deviation for a given $\Delta_6$ mass and neglected all other interactions.
We see in particular that the $\Delta_6$ can be narrow for low enough masses, while for masses of the order of TeV, it becomes very broad and thus more difficult to isolate in spectra. 

\begin{figure}[th]
\begin{center}
\includegraphics[width=9cm]{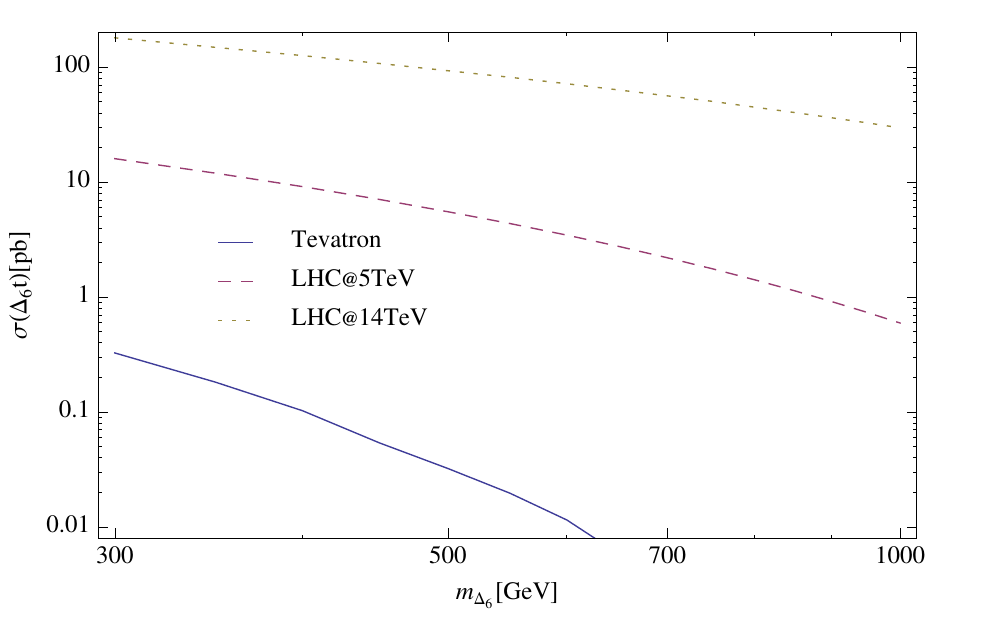}
\end{center}
\caption{ \label{fig:dt} Dependence of the hadronic production cross sections of  $t+\Delta_6$ at the Tevatron and the LHC (for $\sqrt s = 5,\, 14$ TeV) on the mass of ${\Delta_6}$.  Values of $g_{6}^{ut}$ are chosen according to eq. (\ref{eq:fit}).}
\end{figure}

Perhaps the most promising search strategy would be to study the spectrum of the $t\bar t+j$ production and search for (narrow) resonances in the invariant mass of the light jet together with a top or an anti-top ($m_{tj}$ and $m_{\bar t j}$). The cross sections for the production of $t+\Delta_6^*$ (or $\bar t + \Delta_6$) pairs at the Tevatron and the LHC are shown in figure \ref{fig:dt}.
We see that the production cross section at the LHC is at the order of magnitude of the total SM $t\bar t$ cross section over  the complete interesting mass range for the $\Delta_6$ which makes this channel indeed very prospective.

\begin{figure}[th]
\begin{center}
\includegraphics[width=9cm]{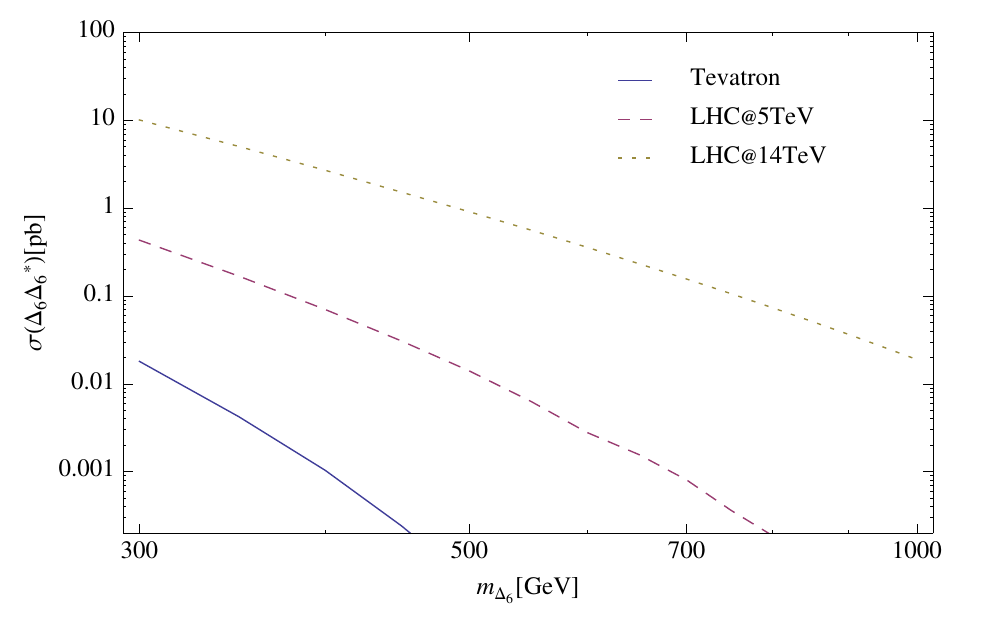}
\end{center}
\caption{ \label{fig:6} Dependence of the hadronic production cross sections of  $\Delta_6^* \Delta_6$ at the Tevatron and the LHC (for $\sqrt s = 5,\, 14$ TeV) on the mass of ${\Delta_6}$.}
\end{figure}

The $\Delta_6$ can also be pair produced in hadronic collisions, decaying into $t\bar t$ pair plus two jets. The production process proceeds though the QCD couplings of the $\Delta_6$, so the $\Delta_6 \Delta^*_6$ cross section only depends on  the $\Delta_6$ mass. At the partonic level, the $\sigma(q\bar q \to \Delta_6 \Delta_6^*)$ and $\sigma(gg \to \Delta_6 \Delta_6^*)$ read \cite{Chen:2008hh}
\begin{eqnarray}
\sigma(q\bar q \to \Delta_6 \Delta_6^*) &=& \frac{2\pi \alpha_s^2}{27 \hat s} \hat \beta_\Delta^3\\
\sigma(g g \to \Delta_6 \Delta_6^*) &=&   \nonumber\\
&&\hspace{-2cm} 
\frac{\pi \alpha_s^2}{96 \hat s} \left\{3 \hat \beta_\Delta  \left(3-5 \hat \beta_\Delta ^2\right)
-16 \hat \beta_\Delta  \left(\hat \beta_\Delta ^2-2\right) \right. \nonumber\\
&&\hspace{-2.9cm} \left.
+\left[8 \left(\hat \beta_\Delta ^4-1\right)-9 \left(\hat \beta_\Delta ^2-1\right)^2\right] \log \left|\frac{\hat \beta_\Delta +1}{\hat \beta_\Delta -1}\right|\right\}\,,
\end{eqnarray}
where $\hat \beta_\Delta = \sqrt{1-4m_{\Delta_6}^2 / \hat s}$. The resulting cross sections at the Tevatron and the LHC are shown in figure \ref{fig:6}.
We see, that the production cross section at the Tevatron only becomes sizable for very low $\Delta_6$ masses. This channel may be more prospective at the LHC, where the cross sections of the order of $1$ pb can be expected even for $\Delta_6$ masses around $0.5$ TeV. The search may be aided by the fact, that the $\Delta_6$'s would appear as resonances in the invariant masses of a top quark and one of the hard jets.


\section{Summary}

The interplay of nonsupersymmetric $SU(5)$ GUT model  which contains the $45$-dimensional  representation with light scalar states and the $ t \bar t $ hadronic production phenomenology results in very interesting consequences.


Within the $SU(5)$ GUT model we found  correlations between the masses of $\Delta_6$ and $\Delta_1$ and the partial proton decay width. In the particular scenario of a TeV scale $\Delta_6$, present constraints also require a TeV scale $\Delta_1$ as well.  A moderate increase in the precision measurement of the $p \to e^+ \pi^0$ decay width will constrain presently allowed ranges of the light scalar  masses.


The contribution of $\Delta_6$ to the production of $ t \bar t $ at the Tevatron does not spoil the successful standard model prediction for the total hadronic cross section, whereas  the associated forward-backward asymmetry can be enhanced and account for the experimental result provided the $\Delta_6$ mass is above 300 GeV.  On the other hand, some tension is already present when comparing with the high $m_{t\bar t}$ spectrum. The constraint gets stronger with larger $\Delta_6$ mass, disfavouring masses above TeV. On the other hand, sizable $\Delta_1$ contributions cannot accommodate experimental results, constraining the mass and couplings of this state. These conclusions are model independent and apply to any scenario with light scalar colored triplets and/or octets.




Some implications of the light $\Delta_6$ scalar in the ongoing and future experiments have been discussed. 
The best strategy for the experimental search for this state would be to study the spectrum of the $ t \bar  t $+jet production and search for resonances in the invariant mass of the light jet together with a top or an anti-top. The $\Delta_6$  can also be pair produced in hadronic collisions resulting in $t\bar t+$2 jets final states. This channel seems to be more promising at the LHC than at the Tevatron.


\section{Note added}

During the completion of this work, refs. \cite{Shu:2009xf, Arhrib:2009hu} appeared, where the effects of colored scalars on the FBA are also considered. Compared to these works, we not only study the influence
of colored scalars on the FBA in a model independent way but also consider a very concrete and viable scenario, where phenomenological constraints require more than one colored scalar to be light. In addition, we study several existing and prospective future experimental constraints or signatures of $\Delta_6$, some of which have not been considered in the afore mentioned papers. In the model independent parts of
our work we confirm results of \cite{Shu:2009xf} and the updated version of \cite{Arhrib:2009hu}.

\begin{acknowledgments}
I. D. would like to thank Goran Senjanovi\'{c} for fruitful discussions. 
  This work is supported in part by the European
  Commission RTN network, Contract No. MRTN-CT-2006-035482
  (FLAVIAnet), and by the Slovenian Research Agency.
\end{acknowledgments}

\appendix

\section{$u g \to \bar t \Delta_6$ amplitudes}

Associated production of $\Delta_6$ in a hadron collider $u g \to \bar
t \Delta_6$ has three contributing diagrams at tree-level (fig.~8).
To avoid fermion number flowing into the $\Delta_6 \bar{u}\bar{t}$
vertex, one can work with the $t^c$ field to make fermion lines
going through the diagram and also correspondingly change the QCD
vertex $g \overline{t^c} t^c$. For color indices arrangement $u_a g^A
\to \bar t_b \Delta_{6c}$ the respective contributions of the three
diagrams of fig.~8 are
\begin{subequations}
\begin{align}
 a_1 &= i g_6^{ut*}\, \epsilon_{dbc}\, \left(\overline{t^c} P_R
   \frac{i(\slashed{p}_{\bar t}+\slashed{p}_{\Delta_6})}{s} [ig\gamma^\mu T^A_{da}] u\right)\, \epsilon^A_\mu,\\
 a_2 &= i g_6^{ut*}\, \epsilon_{a dc}\, \left(\overline{t^c} [-ig\gamma^\mu T^A_{db}] \frac{i(\slashed{p}_u-\slashed{p}_{\Delta_6}+m_t)}{u-m_t^2} P_R u\right)\, \epsilon^A_\mu,\\
 a_3 &= i g_6^{ut*}\, \epsilon_{abd}\, \left(\overline{t^c} P_R
   u\right) \frac{i}{t-m_{\Delta_6}^2} [i g T^A_{cd}
 (p_u-p_t+p_\Delta)^\mu]\,\epsilon^A_\mu.
\end{align}
\end{subequations}
The resulting partonic differential cross section, averaged over polarizations and
colors of the scattered and final state particles is
\begin{equation}
 \frac{d\sigma_{u g \to \bar t \Delta_6}}{dt} = \frac{\alpha_s |g_6^{ut}|^2}{48 s^2} \left|\mathcal{A}_1 + \mathcal{A}_2+\mathcal{A}_3\right|^2\,,\vspace{1cm}
\end{equation}
where individual diagonal and interference terms are
\begin{align}
|\mathcal{A}_1|^2 = & \frac{m_t^2-u}{s}\\
|\mathcal{A}_2|^2 = & \frac{(t-m_t^2) (m_{\Delta_6}^2+t)}{(t-m_{\Delta_6}^2)^2}\\
|\mathcal{A}_3|^2 = & \frac{m_t^4+3 m_t^2 (m_{\Delta_6}^2-u)-m_t^2 t-s u}{(u-m_t^2)^2}\\
2\mathrm{Re}(\mathcal{A}_1 \mathcal{A}_2^*) =& -\frac{2 m_{\Delta_6}^2 (m_t^2-t) - 2 m_t^4 + m_t^2(s+2t)+s t}{s(m_{\Delta_6}^2-t)}\\
2\mathrm{Re}(\mathcal{A}_2 \mathcal{A}_3^*) =& \frac{(m_t^2-t)(m_t^2+m_{\Delta_6}^2)-(m_t^2+t)(m_{\Delta_6}^2-u)}{4(u-m_t^2)(t-m_{\Delta_6}^2)}\\
2\mathrm{Re}(\mathcal{A}_3 \mathcal{A}_1^*) =& 2 \frac{m_t^2(s-u)-m_{\Delta_6}^2 t + 2 m_{\Delta_6}^2 m_t^2 -u s}{s(u-m_t^2)}\,.
\end{align}
Mandelstam variables we used here are $t=(p_u -p_{\bar t})^2$ and
$u=(p_u-p_{\Delta_6})^2$.

\end{document}